\documentstyle[aps,prl,twocolumn,epsf]{revtex}
\draft

\begin{document}
\twocolumn[\hsize\textwidth\columnwidth\hsize\csname 
@twocolumnfalse\endcsname

\title{
Statistical Theory of Energy Transfer to
Small and Chaotic Quantum Systems 
Induced by a Slowly-Varying External Field
}
\author{
Yasuhiro Higashiyama  
and Akira Shimizu\cite{ca}
}
\address{
Department of Basic Science,
University of Tokyo,
3--8--1 Komaba, Meguro-ku, Tokyo 153--8902, Japan
\\
and
Core Research for Evolutional Science and Technology (CREST), JST
}
\date{\today}
\maketitle

\begin{abstract}
We study nonequilibrium properties of 
small and chaotic quantum systems,
i.e., non-integrable systems whose size 
is small in the sense that 
the separations of energy levels 
are non-negligible as compared with 
other relevant energy scales.
The energy change $\Delta E$ 
induced by a slowly-varying external field $\lambda(t)$
is evaluated 
when the range of the variation $\Delta \lambda$ is large
so that the linear response theory breaks down.
A new statistical theory is presented, 
by which we can predict 
$\langle \Delta E \rangle$, 
the average of $\Delta E$ over a finite energy resolution $\delta E$, 
as a function of $\lambda(t)$
if we are given 
the density of states smeared over $\delta E$, 
the average distance of the anticrossings, 
and a constant $K \lesssim 1$.
\end{abstract}

\pacs{PACS numbers: 05.90.+m, 05.45.-a, 05.30.-d, 05.70.Ln}
]

Recently, active research has been devoted to 
small and chaotic quantum (SCQ) systems,
i.e., non-integrable systems whose size 
is small in the sense that 
the separations of energy levels 
are non-negligible as compared with 
other relevant energy scales \cite{review}.
Such systems are frequently encountered in many fields
of physics \cite{review}, 
such as excited states of molecules \cite{molecules} and
those of quantum dots \cite{QD}.
%
SCQ systems have the universal properties 
that level spacings obey a universal distribution,
and that the energy levels exhibit anticrossings 
as a function of an external parameter $\lambda$ \cite{review}.
Note that these are {\em static} properties.
In contrast, 
only limited knowledge has been obtained 
for {\em dynamical} or {\em nonequilibrium} properties
of SCQ systems
\cite{previous_dynamics}.
A major difficulty is that 
solving the time-dependent 
Schr\"odinger equation (TDSE) becomes impossible when
$f \gtrsim 4$,
where $f$ is the 
degrees of freedom of the system,
for the most common 
case where the ``coordinate'' of 
each degree of freedom takes continuous values.
This is because the computational time generally 
grows exponentially with $f$.
If $f$ were huge enough, on the other hand,
the nonequilibrium thermodynamics 
would be applicable, 
by which one could predict 
nonequilibrium properties 
much more easily than by solving the TDSE.
For SCQ systems, 
however, the nonequilibrium thermodynamics 
is not applicable because its 
basic assumptions, such as the local 
equilibrium, are not satisfied.
The purpose of the present Letter 
is to propose a new theory, 
called the random-probability netweork (RPN) method,
by which {\em nonequilibrium} properties of 
SCQ systems can be 
predicted easily, without solving the TDSE
\cite{nondynamical}.

We consider an SCQ system
subject to an external field $\lambda(t)$,
which represents any external disturbance, 
such as a magnetic or electric field, and 
the position coordinate of a moving ``wall''
(a portion of another system which couples to the
SCQ system).
%
The energy change $\Delta E$ 
induced by the variation of $\lambda(t)$ \cite{change}
is evaluated 
under the conditions that (i) 
the range of the variation $\Delta \lambda$ is large
so that 
{\em $\Delta \lambda$ cannot be treated as a small perturbation}
and the linear response theory breaks down, 
and that (ii) the typical time scale 
$T$ of the variation of $\lambda$ is long
(so that $r$ of Eq.\ (\ref{r}) $\ll 1$).
%
Without solving the TDSE, 
we can easily 
predict $\langle \Delta E \rangle$, 
the average of $\Delta E$ over a finite energy resolution $\delta E$, 
%
by the RPN method if we are given a constant $K$ ($\lesssim 1$) 
and two ``fundamental functions'' which contain only small information
of the 
system.

To be concrete, we explain the RPN method using a simple model.
However, since the basic idea is based on the universal properties
mentioned above,
we expect that the RPN method is generally 
applicable to SCQ systems.
Consider the time evolution of 
coupled rotators (angular momenta $\hat {\bf L}$ and $\hat {\bf M}$),
whose Hamiltonian is \cite{this_model}
\begin{equation}
\hat{\cal H}(t) 
= 
\lambda(t)\left(\hat L_{z}+ \hat M_{z}\right) + \hat L_{x} \hat M_{x}
\equiv \hat{\cal H}_{1}(t) + \hat{\cal H}_{2},
\label{Hamiltonian}
\end{equation}
in an appropriate unit system, 
in which the energy, $\lambda$ and $\hbar$ 
are dimensionless \cite{this_model}.
The external field $\lambda(t)$ is taken as
$
\lambda(t) = 
1 - \Delta \lambda [1-(t/T)^2]^4
$
for $-T \le t \le T$
and 
$
\lambda(t) = 1
$
for $|t| > T$ \cite{singularity}.
Since we are interested in the case where 
$\Delta \lambda$ is large, we here take $\Delta \lambda = 0.5$.
For this $\lambda(t)$, 
$\hat {\cal H}$ takes the same form at 
both ends, 
$t=\pm T$,
and 
$\Delta E$ can be finite 
only when transitions occur between different levels. 
However, it should be stressed that 
the RPN method is also 
applicable to the case of $\hat {\cal H}(T) \neq \hat {\cal H}(-T)$, 
for which 
$\langle \Delta E \rangle$ 
consists of both the shifts of energy levels and 
the transitions between them.

\begin{figure}[htbp]
\epsfxsize = 8cm
\centerline{
\epsfbox{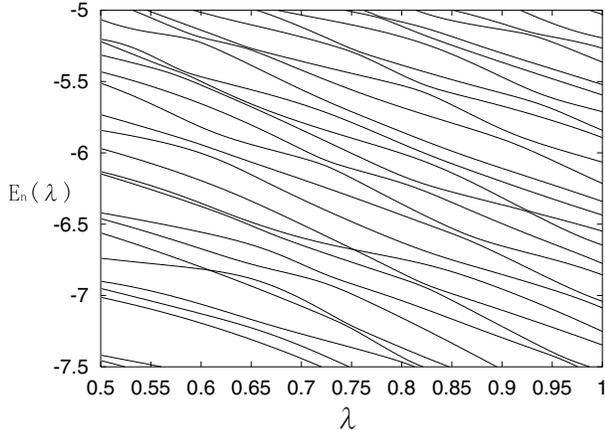}
}
\medskip

\caption{
The energy levels as a function of $\lambda$.
}
\label{fig_espec}
\end{figure}

\begin{figure}
\epsfxsize = 8cm
\centerline{\epsfbox{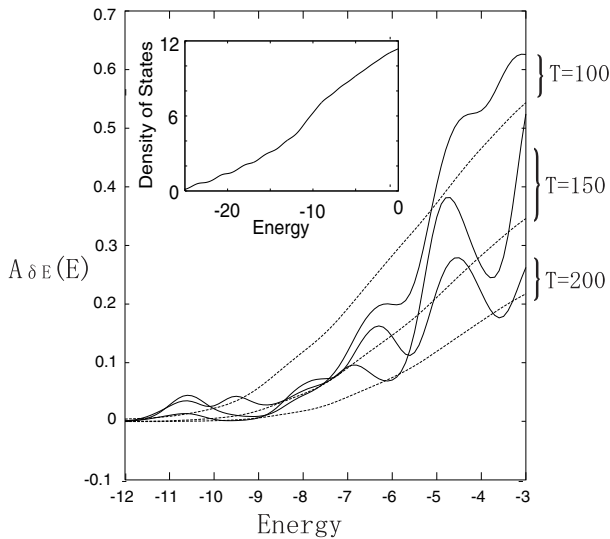}}
\medskip

\caption{
The integrated energy change $A_{\delta E}(E)$ for 
$T=$ 100, 150, 200,
obtained by solving the TDSE (solid lines) and
by the RPN method (dashed lines).
Inset: 
The 
smeared density of states $\rho_{\delta E}(E;\lambda)$
for $\lambda=1.0$, $\delta E =1.0$.
}
\label{fig_A}\end{figure}

As in the case of the statistical mechanics, 
{\em we must first classify quantum states according to 
constants of motion}, to obtain
a set of subspaces, each of which 
has no constant of motion.
{\em It is in each subspace that the system has chaotic natures 
and the RPN method is applicable}.
We here consider the subspace of
$J_{z}/\hbar= even$, $l=m=15$, $\Sigma = +$.
Here,
$J_{z} \equiv L_{z}+M_{z}$,
$\hbar^{2}l(l+1)$ and $\hbar^{2}m(m+1)$
are the quantized values of $\hat L^2$ and $\hat M^2$,
and $\Sigma$ denotes
the parity under the exchange
of $L_z$ and $M_z$.
We take $\hbar=0.322748612$, 
so that the corresponding classical system is chaotic \cite{this_model}.
The eigenvalues $E_n(\lambda)$ 
exhibit anticrossings as functions of $\lambda$
as shown in Fig.\ \ref{fig_espec}.
%
In the inset of Fig.\ \ref{fig_A}, we plot 
the density of states 
$\rho_{\delta E}(E;\lambda)$ 
which is smeared over a finite energy range $\delta E$.
Here and after, we consider the $E<0$ part only, because
the system is symmetric about $E=0$.
On an average, $\rho_{\delta E}(E;\lambda)$
is an increasing function of 
$E$ for $E < 0$ (whereas it is decreasing for $E>0$),
and the average energy spacing ($= 1/ \rho_{\delta E}$) 
decreases with $E$.
We have confirmed that 
the normalized spacing, 
$s_n(\lambda) \equiv 
\rho_{\delta E}(E;\lambda)(E_{n+1}(\lambda)- E_n(\lambda))$, 
obeys the  Wigner distribution,
$ 
P(s) \simeq (\pi s/2) \exp(-\pi s^2/4)
$ \cite{review}.

\begin{figure}[htbp]
\epsfxsize = 8cm
\centerline{\epsfbox{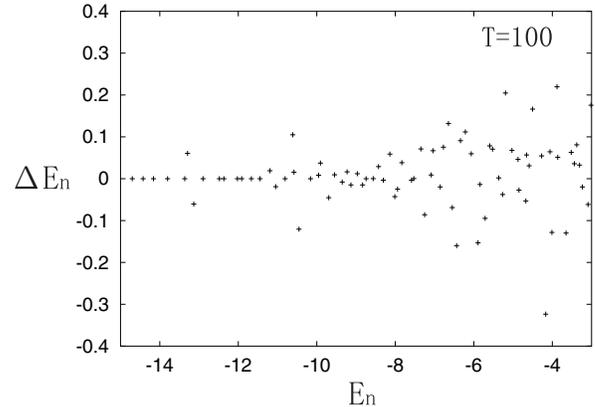}}
\medskip

\caption{
The energy change
for $T=100$,
as a function of the initial energy $E_n$.
}
\label{deltaE}
\end{figure}

Before presenting the RPN method,
we first explore dynamical properties of the system by 
numerically solving 
the TDSE,
$ 
\imath \hbar \frac{\partial}{\partial t}|\psi(t)\rangle = \hat{\cal H}(t)|\psi(t)\rangle.
$ 
Here, we use the split operator method \cite{SOM}
to obtain the time evolution operator 
$\hat U(t_2,t_1)$ defined by 
$ 
|\psi(t_2)\rangle = \hat{U}(t_2,t_1)|\psi(t_1)\rangle.
$
Noting that $\hat {\cal H}(t)$ is constant for $|t| \geq T$, 
we take one of its eigenstates $|\psi_n\rangle$,
where $
\hat {\cal H}(|t| \geq T) |\psi_n\rangle = E_n |\psi_n\rangle
$,
as the initial state.
The state at $t \geq -T$ is denoted by 
$
|\psi_{[n]}(t)\rangle 
\equiv 
\hat{U}(t,-T) |\psi_n\rangle.
$
We are interested in 
the energy change $\Delta E_n$ between the initial 
($t=-T$) and the final ($t=T$) states \cite{change};
$ 
\Delta E_n 
\equiv 
\langle \psi_{[n]}(T)| \hat{\cal H}(T) | \psi_{[n]} (T) \rangle
-E_n.
$ 
This is plotted against $E_n$ for $T=100$
in Fig.\ \ref{deltaE}.
It is seen that 
$\Delta E_n$ scatters rather randomly as a function of $E_n$.
It is therefore {\em impossible} to predict 
$\Delta E_n$ of {\em individual} levels, 
without solving the TDSE.

To see what predictions are possible and meaningful,
let us consider typical experiments on SCQ systems.
Suppose that one tries to prepare the initial state 
in $|\psi_n\rangle$.
However, except for the ground state, 
it is difficult to prepare the initial state 
of SCQ systems accurately.
Hence, {\em one can actually prepare the initial state 
in a finite energy range}
$E_n - \delta E/2 \leq E \leq E_n + \delta E/2$, 
where $\delta E$ denotes the energy resolution of the
apparatus.
Therefore, the initial state should be 
an incoherent mixture or a coherent superposition
of $|\psi_m\rangle$'s of 
$|E_m - E_n| \lesssim \delta E/2$.
Let $\Delta E_{\delta E}^{(j)}(E)$ be the measured value of the
energy change in the $j$th one of such experiments, where
$E=E_n$.
{\em This quantity fluctuates from experiment to experiment} 
for two reasons:
One is the random fluctuation of $\Delta E_n$ 
as a function of $E_n$ mentioned above; 
this fluctuation appears in $\Delta E_{\delta E}^{(j)}(E)$
because the intial state is prepared randomly in 
the energy range $|E_m - E| \lesssim \delta E/2$. 
The other reason is the quantum-mechanical fluctuation;
even if $\delta E \to 0$,
only the {\em average} of $\Delta E_{\delta E}^{(j)}(E)$ 
{\em over many experiments} would agree with $\Delta E_n$.
Namely, even for the same initial state $|\psi_n\rangle$,
the measured value of {\em each} experiment scatters about $\Delta E_n$,
with the standard deviation,
$ 
D_n 
\equiv
[
\langle \psi_{[n]}(T)| 
(
\hat{\cal H}(T)-E_{n}
)^2
| \psi_{[n]} (T) \rangle
-(\Delta E_n)^2
]^{1/2}.
$ 
By numerically computing $D_n$,  
we find that $D_n \sim \Delta E_n$.
Because of these large fluctuations, 
{\em a single experiment is insufficient} for comparison 
of the experimental result with a theoretical one, for an SCQ system.
We must therefore perform many experiments, and 
we are most interested in the {\em average} 
$ 
\Delta E_{\delta E}(E) 
$
of their results.
We will argue later, and have confirmed numerically, 
that the quantum coherence of the 
initial state is irrelevant to this quantity,
under our assumption that {\em $\Delta \lambda$ is large
so that the system experiences many anticrossings
during the variation of $\lambda$}.
Therefore, $\Delta E_{\delta E}(E)$ equals the average
of $\Delta E_m$'s over $m$'s which satisfies 
$|E_m - E| \lesssim \delta E/2$;
\begin{equation}
\Delta E_{\delta E}(E)
\simeq
\frac{\displaystyle{
\int_{E-\delta E/2}^{E+\delta E/2}d\epsilon
\sum_{m}\Delta E_{m}\phi_{\delta E}(\epsilon - E_{m})
}}
{\displaystyle{
\int_{E-\delta E/2}^{E+\delta E/2}d\epsilon
\sum_{m}\phi_{\delta E}(\epsilon - E_{m})
}},
\label{DeltaEd}\end{equation}
where $\phi_{\delta E}(\epsilon)$ is the normalized 
gaussian function with width $\delta E$.
To find statistical tendencies of $\Delta E_{\delta E}(E)$,  
we investigate the {\em integrated energy change} defined by
\begin{equation}
A_{\delta E}(E)\equiv\int_{-\infty}^{E}d\epsilon\cdot 
\sum_{n}\Delta E_{n}\phi_{\delta E}(\epsilon -E_{n}).
\label{A}\end{equation}
We can obtain $\Delta E_{\delta E}(E)$ from 
$A_{\delta E}(E)$ as
\begin{equation}
\frac{
A_{\delta E}(E+\delta E/2)-A_{\delta E}(E-\delta E/2)
}
{
\rho_{\delta E}(E)\delta E
}
=\Delta E_{\delta E}(E).
\label{A2DE}\end{equation}
We plot $A_{\delta E}(E)$ for $\delta E=1.0$,
as a function of $E$, by the solid lines in 
Fig.\ \ref{fig_A}.
We find that $A_{\delta E}(E)$ 
is a smooth and, on an average, increasing 
function of $E$.
It strongly suggests that the prediction about
$A_{\delta E}(E)$ may be possible by a simple theory,
without solving the TDSE.

\begin{figure}[htbp]
\epsfxsize = 8cm
\centerline{\epsfbox{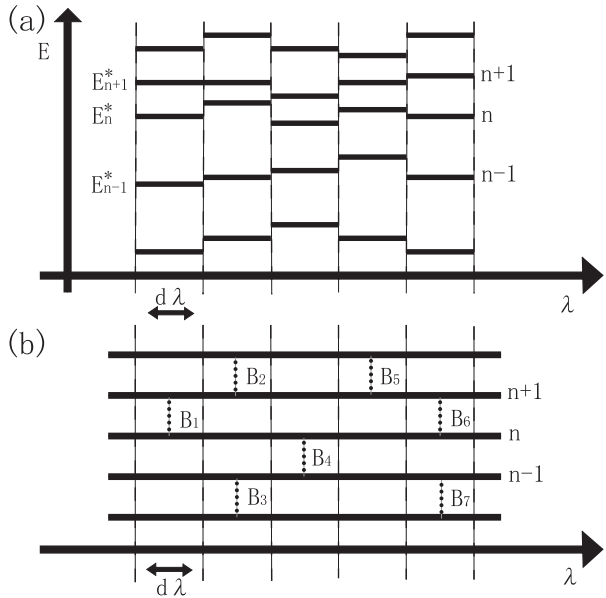}}
\medskip

\caption{
A pair of the RPN diagrams (a) and (b).
}
\label{fig_RPN}
\end{figure}

As such a theory we now present the RPN method, 
by which we can calculate $A_{\delta E}(E)$ 
(hence $\langle \Delta E \rangle$)
as a function of $\lambda(t)$, 
if we are given 
a constant $K$ and two ``fundamental functions,''
which are $\rho_{\delta E}(E; \lambda)$ 
and 
the average distance $\lambda_s(E; \lambda)$
(along the $\lambda$ axis of Fig.\ \ref{fig_espec})
of the anticrossings \cite{average}.
Note that these functions contain  only small 
information on the system
because they represent the distribution of energy levels only, and 
further loss of information occurs by
averaging it over a finite energy $\delta E$.
The RPN method enables us to predict
$A_{\delta E}(E)$ from such small and reduced information.

The RPN method is applicable under the conditions that
(i) $\Delta \lambda$ is large so that 
each level encounters many anticrossings
from $t=-T$ to $T$, and that 
(ii) $T$ is long 
so that most transitions between different levels occur
only at anticrossings, and 
the transition at each anticrossing 
occurs only between adjacent levels.
(This is satisfied if $r$ of Eq.\ (\ref{r}) $\ll 1$.)
Both conditions are
satisfied for $\Delta \lambda = 0.5$ and $T \gtrsim 100$ 
for the present model.
Under these conditions, 
anticrsossings 
may be regarded 
as ``bridges,'' only through which 
transitions occur, and 
the RPN method is formulated as follows.

\noindent{\bf Step 1:} 
Associate the level diagram of the 
original system, Fig.\ \ref{fig_espec}, with many pairs of hypothetical 
diagrams, which we call the RPN diagrams.
Each pair consists of two diagrams (a) and (b), 
as shown in Fig.\ \ref{fig_RPN}.
Diagram (a) is a set of hypothetical energy levels 
$E_1^*, E_2^*, \cdots, E_N^*$, 
where the number $N$ of the levels are taken equal to or 
larger than 
the number of relevant energy levels.
In each small interval 
$[\lambda, \lambda+d\lambda)$,
the hypothetical levels are 
generated randomly according to 
$\rho_{\delta E}(E;\lambda)$ of the original system and 
the Wigner distribution of the spacings.
Here, the small interval $d \lambda$ is taken in such a way that 
$ 
d \lambda
/
\lambda_s(E_n^*(\lambda);\lambda)
\ll
1.
$ 
%
Diagram (b) is a set of straight lines
(labeled by $n= 1, 2, \cdots$) with 
``bridges'' (labeled by $B_1, B_2, \cdots$).
The bridges are randomly generated in such a way 
that the probability of finding one bridge 
in each small interval $[\lambda, \lambda+d\lambda)$
on line $n$ is given by 
$
d \lambda
/
\lambda_s(E_n^*(\lambda);\lambda)
$, 
where $E_n^*(\lambda)$ is given by diagram (a). 
We also impose the condition that 
any two bridges are not on a single line in the same interval.
This constriction is consistent with 
the assumption 
$
d \lambda
/
\lambda_s
\ll
1,
$
which states that 
the probability of finding such two bridges 
in the original system is negligible.

\noindent{\bf Step 2:} 
Attach an $N \times N$ matrix $\mbox{\bf w}$ to each bridge,
where ${\bf w}(B)$ 
at a bridge $B$ between lines $n$ and $n+1$ 
is given by 
$
w_{n,n+1} = w_{n+1,n} = r,
$
$
w_{n,n} = w_{n+1,n+1} = 1-r,
$
and
$
w_{i,j} = w_{j,i} = \delta_{i,j}
$
for $i\ne n,n+1$.
Here, 
$
r \equiv r((E_n^* + E_{n+1}^*)/2;\lambda)
$
represents the transition probability at the bridge, 
where
\begin{equation}
r(E;\lambda)
\equiv
\exp \left[
-
{2 \pi K^2}
{\lambda_s(E;\lambda)}
/
{\hbar |\dot{\lambda}}| {\rho_{\delta E}(E;\lambda)}
\right].
\label{r}\end{equation}
Here, $K$ is a constant ($\lesssim 1$), 
which is considered as a fundamental parameter of the RPN method.
This form of $r(E;\lambda)$ is 
deduced from the Landau-Zener formula for two levels;
$
r = 
\exp ( 
-{2 \pi V^2}/{\hbar |\dot{\lambda}|}{|F_1-F_2|}
)
$.
Here, 
by replacing the difference 
$
|F_1-F_2|
$
between the slopes of the levels with
$
K_2 / \rho_{\delta E}(E;\lambda) \lambda_s
$,
and the energy gap $V$ at the anticrossing with
$
K_3 / \rho_{\delta E}(E;\lambda)
$,
where 
$K_2 \simeq 1$ and $K_3 \lesssim 1$,
we arrive at Eq.\ (\ref{r})
with 
$
K = K_3 / \sqrt{K_2} \lesssim 1
$.

\noindent{\bf Step 3:} 
Construct an $N \times N$ matrix $\mbox{\bf W}$ 
as follows.
Draw the vertical line at 
$\lambda = \lambda(-T)$ 
in diagram (b).
Associate the vertical line with the matrix $\mbox{\bf W}$, 
and take it as the unit matrix.
Move the vertical line as a function of $t$, 
according to $\lambda = \lambda(t)$.
When the vertical line 
encounters a bridge $B$, multiply ${\bf W}$ 
by ${\bf w}(B)$ of the bridge:
$ 
{\bf W} \to  {\bf w}(B) {\bf W}.
$
If the line encounters two or more bridges simultaneously,  
multiply ${\bf W}$ by ${\bf w}$'s of the bridges,  
where, the order of ${\bf w}$'s are arbitrary 
because such ${\bf w}$'s of the bridges {\em on the 
same vertical line} commute with each other.
In this way, ${\bf W}$ at the final time ($t=T$) becomes 
the product of ${\bf w}(B)'s$ along the ``path'' of
$\lambda(t)$.
We interpret its element $W_{mn}$ as the 
probability of the transition 
from line $n$ at $t=-T$ to line $m$ at $t=T$,
of the random probability network.

\noindent{\bf Step 4:}
Evaluate the average change of the hypothetical energy by
$ 
\Delta E_n^*
\equiv
\sum_m
(E_m^*(\lambda(T)) - E_n^*(\lambda(-T)) W_{mn},
$ 
which is the counterpart of $\Delta E_n$ of the original 
system.
Calculate the {\em integrated energy change of the RPN} by
\begin{equation}
A_{\delta E}^*(E)\equiv\int_{-\infty}^{E}d\epsilon\cdot 
\sum_{n}\Delta E_{n}^* \phi_{\delta E}(\epsilon -E_{n}^*).
\label{ARPN}\end{equation}

\noindent{\bf Step 5:}
Repeat steps 1-4 many times, to obtain many 
$A_{\delta E}^*(E)$'s, which correspond to
different pairs of the RPN diagrams.
Calculate the average of $A_{\delta E}^*(E)$'s.
This average is the result for $A_{\delta E}(E)$ by  
the RPN method.

The total computational time is tremendously 
reduced as compared with solving the TDSE; 
{\it e.g.}, from a few days to a few seconds
for the model of Eq.\ (\ref{Hamiltonian}).
The reduction becomes much more drastic as $f$ is increased.

To check the validity of the RPN method, 
we compare $A_{\delta E}(E)$ of the RPN method 
with that obtained by solving the TDSE.
In the RPN method we use 
$\rho_{\delta E}(E;\lambda)$ and $\lambda_s(E;\lambda)$ 
that are obtained by solving the 
time-{\em independent} Schr\"odinger equation.
The constant $K$ is determined so as to optimize 
the agreement between the results of the RPN method and 
the TDSE. 
(Hence, $K$ is the only free parameter in the comparison.)
This gives $K=0.177$, in consistent with $K \lesssim 1$.
The dashed lines in Fig.\ \ref{fig_A} plot 
$A_{\delta E}(E)$ obtained by the RPN method
for various values of $T$.
On an average, they agree with the solid lines
obtained by solving the TDSE, 
although
the fine structures, peaks and dips, of the solid lines are
absent in the dashed lines.
It is expected that such peaks and dips would be shallower for 
larger systems.
Considering also the tremendous reduction of the computaional time, 
the overall agreement seems satisfactory.
%
The reason why the RPN method gives good results 
is partly that we have introduced the finite 
energy resolution $\delta E$, which results in 
the {\em average over many initial states}.
This smears out strong dependence on the initial state.
Another reason is that we have assumed a large value of $\Delta \lambda$. 
As a result, {\em the wavefunction experiences many
anticrossings} under the variation of $\lambda(t)$, 
and effects of the phase coherence on $\langle \Delta E \rangle$ 
are almost destroyed on an average.
Detailed discussions on this point will be described elsewhere.

As we have done above, we may obtain the fundamental functions 
by solving the time-{\em independent} 
Schr\"odinger equation, 
which is much easier than solving the TDSE.
For larger systems, or for systems whose Hamiltonian is unknown, 
the fundamental functions and $K$
may be obtained from some
experiments (e.g., on the specific heat) 
and/or experiences (in similar systems), 
as in the case of estimating thermodynamical functions 
from experiments.
To develop a systematic method
to perform this program is a subject of future study.

We finally note that $A_{\delta E}(E)$ is, 
on an average, an increasing function of $E$.
This means that {\em if $\lambda(t)$ is varied repeatedly}, 
then, as a result of accumurated transitions, 
the energy of the system tends toward $E=0$,
where $\rho_{\delta E}(E)$ is highest.
We thus find that 
the system tends to ``climb'' the curve of
the smeared density of state.
Therefore, an SCQ system tends to absorb the energy if 
its $\rho_{\delta E}(E)$ is an increasing function of $E$.


\begin{references}
\bibitem{change}
The energy change equals the energy transfer 
between the SCQ system and the external system that causes the variation 
of $\lambda(t)$.
\bibitem{singularity}
Some models, such as kicked rotors, have 
singularies in the $t$ dependence of $\hat {\cal H}(t)$ 
or $\lambda(t)$. However, such singularities should not be 
present in real physical systems.
The RPN method is applicable to the latter case.
Although $\lambda(t)$ of Eq.\ (\ref{Hamiltonian}) has 
weak singularities at $t=\pm T$, we have confirmed numerically that
their effects are negligible for the values of $T$ assumed in the 
present paper.
\bibitem{average}
The average is taken over $\delta E$ and $\delta \lambda$, 
where $\delta \lambda$ is taken large enough ($=0.1$ here) so that 
$\lambda_s$ is well defined.
\end{references}
\end{document}